\documentclass[twocolumn,superscriptaddress,floatfix,preprintnumbers, nofootinbib,hyperref]{revtex4-2} 
\pdfoutput=1
\usepackage[colorlinks=true,breaklinks=true]{hyperref}
\usepackage[normalem]{ulem}
\usepackage[utf8]{inputenc}
\hypersetup{allcolors=[rgb]{0.0 0.0 0.6},linkcolor=[rgb]{0.75 0.05 0.05}}
\usepackage{amsmath,amssymb}
\usepackage{epsfig}  
\usepackage{graphicx}   
\usepackage{slashed}       
\usepackage{tikz}
\usepackage{tikz-feynman}
\usepackage{url}
\usepackage{color}
\usepackage{multirow}
\usepackage{comment}
\usepackage{amssymb}
\usepackage{orcidlink}

\DeclareMathOperator{\eV}{eV}

\DeclareMathOperator{\MeV}{MeV}

\DeclareMathOperator{\s}{s}

\newcommand{\bk}{{\bf k}}
\newcommand{\bl}{{\bf l}}

\newcommand{\bq}{{\bf q}}
\newcommand{\bp}{{\bf p}}

\allowdisplaybreaks

\bibpunct{[}{]}{,}{n}{}{,}
\definecolor{ForestGreen}{RGB}{34,139,34}

\begin{document}

\title{Axion emission from supernovae: a cheatsheet}

\author{Pierluca Carenza \orcidlink{0000-0002-8410-0345}}\email{pierluca.carenza@fysik.su.se}
\affiliation{The Oskar Klein Centre, Department of Physics, Stockholm University, Stockholm 106 91, Sweden
}

\date{\today}
\smallskip

\begin{abstract}
Supernovae provide fascinating opportunities to study various particles and their interactions. Among these there are neutrinos, axions, and other light weakly interacting particles, which play a significant role in our understanding of fundamental physics. 
In this study, the focus lies on the recent advancements made in characterizing axion emission from nuclear matter within the context of supernovae. The main production mechanisms for axions coupled with nucleons, bremsstrahlung and pion-axion conversion, are extensively discussed. These findings shed light on the behavior of axions in dense and hot nuclear matter, encountered in these extreme astrophysical environments. 
\end{abstract}

\maketitle

\section{Introduction}
\label{sec:1}

In recent years there has been a renewed interest towards light exotic particles, their direct signatures in laboratory searches and indirect evidences in astrophysical phenomena~\cite{Lanfranchi:2020crw,Antel:2023hkf}.
In this picture, supernovae (SNe) are valuable cosmic laboratories for studying feebly-interacting particles lighter than approximately $100$~MeV~\cite{Raffelt:1990yz,Raffelt:1987yt}.
A significant milestone in our understanding of SN physics was achieved through the detection of the neutrino burst associated with the nearby SN 1987A~\cite{Kamiokande-II:1987idp,Hirata:1988ad,Bionta:1987qt}. The dozen of detected neutrino events allowed us to better understand neutrino properties~\cite{Kolb:1987qy,Raffelt:1990yu} and constrain other exotic particles, as axions~\cite{Turner:1987by,Brinkmann:1988vi,Chang:2018rso,Lucente:2022vuo}, dark photons~\cite{DeRocco:2019njg}, gravitons~\cite{Hannestad:2001jv} and scalars mixed with the Higgs boson~\cite{Balaji:2022noj}.
The most important tool used in this context  to constrain new particles is the \emph{cooling argument}~\cite{Raffelt:1990yz}: the energy subtracted to the SN by the emission of exotic particles has to be smaller than the standard losses due to neutrinos, otherwise the expected duration of the neutrino burst would be shorter than the observations of SN 1987A.

This criterion was famously applied to the case of axions, originally introduced to solve the \emph{strong-CP problem}, i.e. the unnatural smallness of the observed CP-violating interactions in Quantum Chromodynamics (QCD)~\cite{Weinberg:1977ma,Wilczek:1977pj}, and good dark matter candidates~\cite{Abbott:1982af,Preskill:1982cy,Dine:1982ah} if their mass is $\sim\mathcal{O}({\rm\mu \eV})$~\cite{Borsanyi:2015cka,Ringwald:2016yge}. 
If axions interact with nucleons, the dense core of SNe is a favorable environment to copiously produce them via nucleon-nucleon bremsstrahlung~\cite{Carena:1988kr,Brinkmann:1988vi,Raffelt:1993ix,Raffelt:1996wa,Carenza:2019pxu} $NN\to NNa$, and pionic Compton-like scatterings (also called pion-axion conversion), $\pi^{-} p\to n a$~\cite{Turner:1991ax,Raffelt:1993ix,Keil:1996ju}.
The cooling argument was initially estimated to set an upper limit on the axion mass at $\sim16$~meV~\cite{Raffelt:2006cw} when neglecting pionic processes (see~\cite{Carenza:2019pxu,Carenza:2020cis,Lella:2022uwi,Lella:2023bfb} for recent developments including the pion-axion conversion).
The role of SNe in determining an upper bound for the axion mass, makes this phenomenology widely studied and relevant to guide experimental searches~\cite{IAXO:2019mpb}.  

For these reasons, we witnessed a fast development in the characterization of the axion emission from the nuclear matter encountered in SNe. First of all, a significant effort was devoted to an accurate calculation of the bremsstrahlung axion emission beyond the One-Pion-Exchange (OPE) approximation, where the interaction between nucleons is assumed to be mediated by a pion~\cite{Turner:1987by,Carena:1988kr,Brinkmann:1988vi}. 
In~\cite{Carenza:2019pxu} it was shown that corrections to OPE include: a non-vanishing mass for the exchanged pion~\cite{Stoica:2009zh}; the contribution from the exchange of two-pions~\cite{Ericson:1988wr}; effective in-medium nucleon masses and multiple nucleon-nucleon scatterings~\cite{Raffelt:1991pw,Janka:1995ir}. Their role is to reduce the axion emissivity of an order of magnitude.
Secondly, in~\cite{Carenza:2020cis} (see also~\cite{Fischer:2021jfm}) it was pointed out that the pion-axion conversion dominates over the bremsstrahlung production in typical SN conditions, given a realistic pion abundance~\cite{Fore:2019wib}. This conclusion was a paradigm shift compared to previous studies, where pionic processes were totally neglected. 

Here, we give a comprehensive and updated description of the axion emission from SNe. In Section~\ref{sec:emissivity} we describe both the axion production via bremsstrahlung and pionic processes. In Section~\ref{sec:luminosity} we describe how to evaluate the axion luminosity from a SN. This quantity is used to set the cooling bound, explained in Section~\ref{sec:cooling}. In Section~\ref{sec:conclusions} we summarize and conclude.

\section{Axion emissivity in nuclear matter}
\label{sec:emissivity}

Axion-nucleon interactions are is described by the following Lagrangian~\cite{DiLuzio:2020wdo,Chang:1993gm}
\begin{equation}
    \begin{split}
        \mathcal{L}_{\rm{int}}&=g_a\frac{\partial_\mu a}{2m_N}\Bigg[C_{ap}\Bar{p}\gamma^\mu\gamma_5p+C_{an}\Bar{n}\gamma^\mu\gamma_5n+\\
        &+\frac{C_{a\pi N}}{f_\pi}(i\pi^+\Bar{p}\gamma^\mu n-i\pi^-\Bar{n}\gamma^\mu p)+\\
        &+C_{aN\Delta}\left(\Bar{p}\,\Delta^+_\mu+\overline{\Delta^+_\mu}\,p+\Bar{n}\,\Delta^0_\mu+\overline{\Delta^0_\mu}\,n\right)\Bigg]\,,
    \end{split}
\label{eq:NuclearInteractions}
\end{equation}
where $g_a=m_N/f_a$ is the dimensionless axion-nucleon coupling suppressed by the Peccei-Quinn scale $f_{a}$, i.e.~the new physics scale introduced by axions, $m_{N}=938$~MeV is the nucleon mass, $C_{aN}$ with $N=p,n$ are model-dependent $\mathcal{O}(1)$ coupling constants, $f_{\pi}=92.4~\MeV$ is the pion decay constant, $C_{a\pi N}=(C_{ap}-C_{an})/\sqrt{2}g_{A}$~\cite{Choi:2021ign} is the axion-pion-nucleon coupling and ${C_{aN\Delta}=-\sqrt{3}/2\,(C_{ap}-C_{an})}$ is the axion-nucleon-$\Delta$ baryon coupling, with $g_{A}\simeq1.28$~\cite{Workman:2022ynf} the axial coupling.
For convenience, we define the axion-proton and axion-neutron coupling as $g_{aN}=g_{a}C_{aN}$ for $N=p,n$. The benchmark axion model we use in this work is the Kim-Shifman-Vainshtein-Zakharov (KSVZ) model~\cite{Kim:1979if,Shifman:1979if}, where $C_{ap}=-0.47$ and $C_{an}=0$~\cite{GrillidiCortona:2015jxo}. We consider the KSVZ as a benchmark model because this is one of the simplest axion models and the couplings with nuclear matter are extremely model-independent. Nevertheless, these couplings have an associated uncertainty related to the determination of isoscalar matrix elements. The axion-neutron coupling is, within these uncertainties, compatible with zero. Thus, we assume $C_{an}=0$.

\begin{figure}[t!]
    \centering
  \begin{tikzpicture}[scale=1.1, transform shape]
\begin{feynman}
\vertex (a) {\(N\)} ;
\vertex [above=of a] (b) {\(a\)};
\vertex [right=of a] (c) {\(C_{aN}\)};
\vertex [right=of c] (d)  {\(N\)};

\diagram* {
(a) -- [fermion] (c),
(c) -- [fermion] (d),
(b) -- [scalar] (c),
};
\end{feynman}
\end{tikzpicture}

\begin{tikzpicture}[scale=1., transform shape]
\begin{feynman}
\vertex (a) {\(p\)} ;
\vertex [above=of a] (b) {\(a\)};
\vertex [right=of a] (c) {\(C_{a\pi N}\)};
\vertex [right=of c] (d) {\(n\)};
\vertex [above=of d] (e) {\(\pi^{+}\)};

\diagram* {
(a) -- [fermion] (c),
(c) -- [fermion] (d),
(b) -- [scalar] (c),
(c) -- [scalar] (e),
};
\end{feynman}
\end{tikzpicture}

 \begin{tikzpicture}[scale=1.1, transform shape]
\begin{feynman}
\vertex (a) {\(N\)} ;
\vertex [above=of a] (b) {\(a\)};
\vertex [right=of a] (c) {\(C_{aN\Delta}\)};
\vertex [right=of c] (d)  {\(\Delta\)};

\diagram* {
(a) -- [fermion] (c),
(c) -- [fermion] (d),
(b) -- [scalar] (c),
};
\end{feynman}
\end{tikzpicture}
    \caption{Feynman diagrams of the vertices in Eq.~\eqref{eq:NuclearInteractions}. {\it Upper panel}: axion-nucleon vertex, where $N=p,n$. {\it Middle panel}: example of axion-pion-nucleon vertex. The vertex with $\pi^{-}$ is also possible, but not with $\pi^{0}$. {\it Lower panel}: axion-nucleon-$\Delta$ baryon vertex, where $N=p,n$. The electric charge of the $\Delta$ baryon is chosen in line with the nucleon one.}
    \label{fig:feynm}
\end{figure}
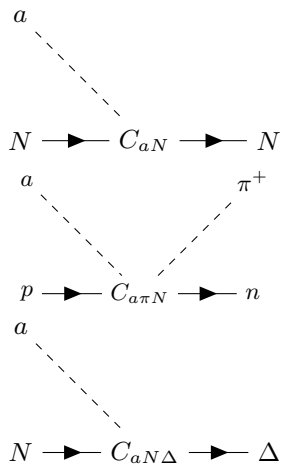

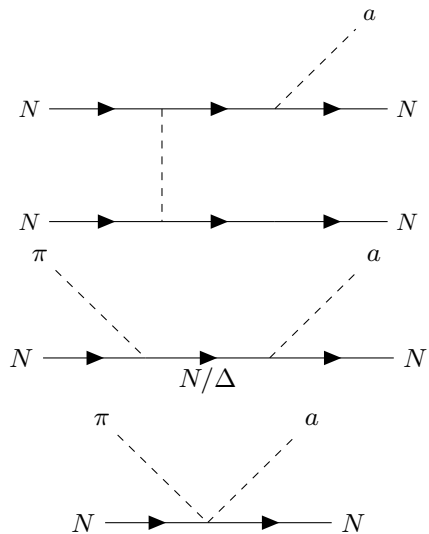
\begin{figure}[t!]
\centering
\begin{tikzpicture}[scale=1., transform shape]
\begin{feynman}
\vertex (a);
\vertex [above=of a] (b);
\vertex [right=of a] (c);
\vertex [left=of a] (d){\(N\)};
\vertex [right=of c] (e) {\(N\)};
\vertex [right=of b] (f);
\vertex [right=of f] (g){\(N\)};
\vertex [left=of b] (h){\(N\)};
\vertex [above right=of f] (k) {\(a\)};

\diagram* {
(d) -- [fermion] (a),
(a) -- [fermion] (c),
(c) -- [fermion] (e),
(h) -- [fermion] (b),
(b) -- [fermion] (f),
(f) -- [fermion] (g),
(f) -- [scalar] (k),
(b) -- [scalar] (a),
};
\end{feynman}
\end{tikzpicture}

 \begin{tikzpicture}[scale=1.1, transform shape]
\begin{feynman}
\vertex (a) {\(N\)} ;
\vertex [right=of a] (b);
\vertex [right=of b] (c);
\vertex [right=of c] (d)  {\(N\)};
\vertex [above left=of b] (e) {\(\pi\)};
\vertex [above right=of c] (f) {\(a\)};

\diagram* {
(a) -- [fermion] (b),
(b) -- [fermion, edge label'=\(N/ \Delta\)] (c),
(c) -- [fermion] (d),
(e) -- [scalar] (b),
(c) -- [scalar] (f),

};
\end{feynman}
\end{tikzpicture}

 \begin{tikzpicture}[scale=1.1, transform shape]
\begin{feynman}
\vertex (a) {\(N\)} ;
\vertex [right=of a] (b);
\vertex [right=of b] (d){\(N\)};
\vertex [above left=of b] (e) {\(\pi\)};
\vertex [above right=of b] (f) {\(a\)};

\diagram* {
(a) -- [fermion] (b),
(b) -- [fermion] (d),
(e) -- [scalar] (b),
(b) -- [scalar] (f),

};
\end{feynman}
\end{tikzpicture}
    \caption{Feynman diagrams of the axion emission processes. Here, $N=p,n$ and the charges of other particles are chosen to satisfy the electric charge conservation at each vertex. {\it Upper panel}: bremsstrahlung of an axion by the interactions of two nucleons, whose effective mediator models a realistic nuclear interaction. {\it Middle panel}: pion-axion conversion process, where the intermediate virtual state can be either a nucleon or a $\Delta$ baryon.  {\it Lower panel}: pion-axion conversion process due to the contact interaction.}
    \label{fig:feynm2}
\end{figure}

In Fig.~\ref{fig:feynm} we show the Feynman diagrams of the vertices appearing in Eq.~\eqref{eq:NuclearInteractions}. The axion-nucleon vertex in the upper panel is the simplest interaction giving rise to axion emission from hot nuclear matter.  
The four-particles interaction vertex shown in the middle panel of Fig.~\ref{fig:feynm} is also known as contact interaction and its impact on the axion emissivity is significant and it was discussed in~\cite{Carena:1988kr,Choi:2021ign}. Recently, \cite{Ho:2022oaw} showed also the importance of the vertex with the $\Delta$ resonance, bottom panel of Fig.~\ref{fig:feynm}, leading to a sizable increase in the axion emissivity.
Therefore, none of these interactions can be neglected in an accurate evaluation of the axion emission from SNe.
In this Section, we explain in details all the ingredients needed to calculate the axion emissivity from a SN via the processes shown in Fig.~\ref{fig:feynm2}.
The Feynman diagram of the bremsstrahlung process is the one in the upper panel, where the interaction between two nucleons is engineered to reproduce nuclear scattering data~\cite{Hanhart:2000ae,Carenza:2019pxu}. The other two diagrams, in the middle and lower panels, contribute to the pion-axion conversion. The former features an intermediate off-shell nucleon state or exciting a $\Delta$-resonance, the latter is possible due to the contact interaction vertex. The interplay between these two processes is non-trivial and it was explored in~\cite{Ho:2022oaw,Lella:2022uwi}.

Regarding pionic processes, we observe that the abundance of positively charged and neutral pions in SNe is expected to be small compared to negatively charged ones~\cite{Fore:2019wib}. Thus, the pion indicated in the diagrams in Fig.~\ref{fig:feynm2} is, for production processes, a negatively charged pion $\pi^{-}$ and these processes read as $\pi^{-}p\to a n$. When the diagrams in Fig.~\ref{fig:feynm2} are read in the opposite way, for axion absorption, more processes are possible and any pion can be included in the diagrams (except the neutral pion for the contact interaction term, lower panel) because the absorption does not depend on pionic abundances. 
In the following, we present a concise description of how to calculate axion emissivities in a SN.

\subsection{Nucleon-nucleon bremsstrahlung}

The spin-summed matrix element of the bremsstrahlung is~\cite{Giannotti:2005tn}
\begin{equation}
    \begin{split}
&S\displaystyle\sum_{\rm{spins}} |\mathcal{M}|^2=\left(
    \frac{{|{\bf p}_a|}}{E_a}\right)^{2}\frac{g_{a}^{2}}{4m_{N}^{2}} E_{a}^{2}\,\bar{M} \,\ ,\\
 &   \bar{M}= \frac{256}{3}m_{N}^{4} E_{a}^{-2} \left(\frac{g_{A}}{2f_{\pi}}\right)^{4}\left(\mathcal{A}_{nn}+\mathcal{A}_{pp}+4 \mathcal{A}_{np}\right)\,,\\
 & \mathcal{A}_{NN}= C_{aN}^{2}
\Bigg[\mathcal{F}_{-}^{2}+\mathcal{F}_{+}^{2}+(1-\xi)\mathcal{F}_{-}\mathcal{F}_{+}\Bigg]\,, \\
&\mathcal{A}_{np}=\left(C_{+}^{2}+C_{-}^{2}\right)\mathcal{F}_{-}^{2}+\left(4C_{+}^{2}+2C_{-}^{2}\right)\mathcal{F}_{+}^{2}+ \\
&-2\left[\left(C_{+}^{2}+C_{-}^{2}\right)-\left(3C_{+}^{2}+C_{-}^{2}\right)\frac{\xi}{3}\right]\mathcal{F}_{-}\mathcal{F}_{+} \,\ ,
\end{split}
    \label{eq:MatElementBrem}
\end{equation}
where in the amplitude $\mathcal{A}_{NN}$ the subscript $NN$ indicates proton-proton and neutron-neutron processes, $C_{\pm}= \left(C_{an}\pm C_{ap}\right)/2$,  $E_{a}$ and $|\bp_{a}|$ are the axion energy and momentum, respectively. We consider massive axions, lighter than nucleons $m_{a}<m_N$. In this case, the only difference with the massless case is the overall axion velocity factor $\beta_a={|{\bf p}_a|}/{E_a}$.
Here, we introduced the following quantities related to the pion propagator, mediating the nucleon-nucleon interaction in the OPE approximation
\begin{equation}
    \begin{split}
    \mathcal{F}_{-}&=\frac{|\bk|^{2}}{|\bk|^{2}+m_{\pi}^{2}}\,,\\
    \mathcal{F}_{+}&=\frac{|\bl|^{2}}{|\bl|^{2}+m_{\pi}^{2}}\,,\\
 \xi&=3\frac{(\bk\cdot\bl)^{2}}{|\bk|^{2}|\bl|^{2}} \, ,\\
\label{eq:CpCn}       
    \end{split}
\end{equation}
where $m_{\pi}$ is the pion mass and $\bk=\bp_{1}-\bp_{3}$ and  $\bl=\bp_{1}-\bp_{4}$ are the exchanged nucleus momenta, labeled as 1 and 2 in the initial state and 3 and 4 in the final state.
In the OPE approximation with massless pions $\mathcal{F}_{+}=\mathcal{F}_{-}=1$. The first two terms in Eq.~\eqref{eq:CpCn} have to be modified to improve the description of nuclear scatterings. Following~\cite{Ericson:1988wr}, the short-range nuclear potential is modeled as an exchange of an effective $\rho$-meson 
\begin{equation}
    \begin{split} \mathcal{F}_{-}&=\frac{|\bk|^{2}}{|\bk|^{2}+m_{\pi}^{2}}-C_{\rho}\frac{|\bk|^{2}}{|\bk|^{2}+m_{\rho}^{2}}\,,\\
    \mathcal{F}_{+}&=\frac{|\bl|^{2}}{|\bl|^{2}+m_{\pi}^{2}}-C_{\rho}\frac{|\bl|^{2}}{|\bl|^{2}+m_{\rho}^{2}}\,,\\
    \end{split}
\end{equation}
where $m_{\rho}=600$~MeV and $C_{\rho}=1.67$.
At this point it is convenient to introduce the center-of-mass $\bp_{0}$ and exchanged momenta variables, $\bp$ and $\bq$, defined as~\cite{Raffelt:1993ix}
\begin{equation}
\bp_{1/2}=\bp_{0}\pm\bp \,, \quad\bp_{3/4}=\bp_{0}\pm\bq\,,\\
\end{equation}
and the angles between them are defined by the following relations
\begin{equation}
\begin{split}
\bp\cdot\bq&=|\bp||\bq|\cos\theta\,,\\
\bp\cdot\bp_{0}&=|\bp||\bp_{0}|\cos\delta\,,\\ \bq\cdot\bp_{0}&=|\bq||\bp_{0}|\left(\sin\delta\sin\theta\cos\phi+\cos\delta\cos\theta\right) \, .\\
\end{split}
\end{equation}
Finally, we introduce the new energy variables
\begin{equation}
\begin{split}
&u=\frac{|\bp|^{2}}{m_{N}T}\,,\quad v=\frac{|\bq|^{2}}{m_{N}T}\,,\quad w=\frac{|\bp_{0}|^{2}}{m_{N}T}\,, \\
\end{split}
\end{equation}
and $ y=m_{\pi}^{2}/m_{N}T$, where $T$ is the temperature of the nucleon plasma in the SN.
In terms of these variables Eq.~\eqref{eq:CpCn} reads
\begin{equation}
    \begin{split}
 \mathcal{F}_\pm &=     \frac{u+v\pm2\sqrt{uv}\cos\theta}{u+v\pm2\sqrt{uv}\cos\theta+y}  \,,   \\
\xi&=3\frac{(u-v)^{2}}{(u+v)^{2}-4uv\cos^{2}\theta} \,.
\label{eq:ffactor}       
    \end{split}
\end{equation}
In order to take into account the $\rho$-meson exchange, we shift the $\mathcal{F}_{\pm}$ terms as
\begin{equation}
    \begin{split}
    &\mathcal{F}_{\pm}\rightarrow \tilde{\mathcal{F}}_{\pm} =\mathcal{F}_{\pm}-C_{\rho}\mathcal{G}_{\pm} \,\ ,\\
&\mathcal{G}_{\pm}=\frac{u+v\pm2\sqrt{uv}\cos\theta}{u+v\pm2\sqrt{uv}\cos\theta+r} \,\ ,\\
&r=\frac{m_{\rho}^{2}}{m_{N}
T} \,\ ,   
    \end{split}
\end{equation}
where $m_{\rho}=600$~MeV and $C_{\rho}=1.67$~\cite{Ericson:1988wr}.
Once that the matrix element for bremsstrahlung is defined, we can calculate the axion spectrum per unit volume as
\begin{equation}
    \begin{split}
        &\left(\frac{d^2n_a}{dE_a\,dt}\right)_{NN}=\int \prod_{i=1}^{4}\left[\frac{2\,d^{3}\bp_{i}}{(2\pi)^{3}2m_{N}}\right]\frac{4\pi E_{a}|\bp_{a}|}{(2\pi)^{3}2E_{a}}\\
    &(2\pi)^{4}\delta^{3}(\bp_{1}+\bp_{2}-\bp_{3}-\bp_{4})    \delta(E_{1}+E_{2}-E_{3}-E_{4}-E_{a})\\
    &S\sum_{\rm spins}|\mathcal{M}|^{2}f_1\,f_2\,(1-f_3)(1-f_{4})=\\
        &=\frac{g_a^2}{16\pi^2} \frac{n_B}{m_N^2}(E_a^2-m_a^2)^{\frac{3}{2}}e^{-\frac{E_a}{T}}  S_\sigma\left(E_{a}\right)\,\Theta(E_a-m_a)\,,
    \label{eq:dnNN}    
    \end{split}
\end{equation}
where $n_{B}$ is the baryon density in the SN core, the Heaviside-$\Theta$ function sets the minimum axion energy to its mass and we integrated the matrix element in Eq.~\eqref{eq:MatElementBrem} over the phase space of the four nucleons. In this analysis we implicitly assume that the axion momentum is negligible compared to the nucleon ones, i.e.~$\bp_{1}+\bp_{2}\simeq\bp_{3}+\bp_{4}$. The four nucleon distributions $f_{i}$ for $i=1,...4$ are Fermi-Dirac distributions in the non-relativistic limit
\begin{equation}
    f(\bp)=\frac{1}{e^{\frac{|\bp|^{2}}{2m_{N}T}-\hat{\mu}}+1}\,,
\end{equation}
where the nucleon degeneracy parameter is
\begin{equation}
    \hat{\mu}=\frac{\mu-m_{N}-U}{T}\,,
\end{equation}
obtained from the chemical potential $\mu$ and nucleon self-energy $U$, including scalar and vector components. When this quantity is positive, nucleons are degenerate, non-degenerate otherwise.
After integrating Eq.~\eqref{eq:dnNN} over the phase space, it is possible to split it into a part dependent on the axion and a term involving only nuclear physics, $S_{\sigma}$, the nucleon structure function~\cite{Raffelt:1996wa}. The structure function is defined as~\cite{Hannestad:1997gc,Raffelt:1993ix}
\begin{equation}
        S_{\sigma}(E_a)=\frac{\Gamma_\sigma}{E_{a}^2+\Gamma^2}s\left(\frac{E_a}{T}\right)\,,
        \label{eq:sigma}
\end{equation}
where the nucleon spin fluctuation rate is
\begin{equation}
\begin{split}
\Gamma_{\sigma}&=4\pi^{-1.5}\rho\left(\frac{g_{A}}{2f_{\pi}}\right)^{4}T^{0.5}m_N^{0.5}\,.
\end{split}
\label{eq:gammasigma}
\end{equation}
In particular, the width $\Gamma$ of the Lorentzian distribution in Eq.~\eqref{eq:sigma} is determined by the following normalization condition~\cite{Raffelt:1996di,Hannestad:1997gc}
\begin{equation}
\begin{split}
&\int_{0}^{\infty} \frac{d E_{a}}{2 \pi} (1+e^{-E_{a}/T})S_{\sigma}(\omega) =\\
&\hspace{1cm}\frac{1}{n_B}
\sum_{i=p,n}\frac{C_{ai}^2 }{C_{ap}^{2}Y_{p}+C_{an}^{2}Y_{n}} \int \frac{2 d^3 {\bf p}}{(2\pi)^3}f_i (1-f_i) \,\ ,
\label{eq:normal}
\end{split}
\end{equation}
where $Y_i$ for $i=p,n$ are the numbers of protons or neutrons per baryon and we assume that the nucleon spins fluctuate in an uncorrelated way.
For a single non-degenerate species, neglecting the Pauli blocking factor 
$(1-f_i)$, the structure function has to be normalized to one. We will work in the ansatz $\Gamma=g\, \Gamma_{\sigma}$, where $g$ is determined by the condition in Eq.~\eqref{eq:normal}.

From this discussion, it is clear that the axion emission does not monotonically increase with $\Gamma_{\sigma}$. At some point it will start decreasing.
Indeed, the axion emissivity is suppressed because nucleons spin-dependent self-interactions create an effectively fluctuating average spin. The stronger the self-interactions, the smaller the averaged effective spin which couples to axions. Therefore, nucleon-nucleon scatterings lead to a suppression of the axion emission~\cite{Raffelt:1996za}. This effect is, in a more general picture, a suppression of the axial interactions between nucleons and any weakly interacting particle, just like neutrinos. If the spin fluctuation rate in Eq.~\eqref{eq:gammasigma} exceeds a few times the temperature, also neutrino-nucleon interactions are strongly suppressed. In~\cite{Keil:1994sm,Janka:1995ir,Sigl:1995ac} it was argued that the spin fluctuation rate should saturate at $\Gamma_{\sigma}\simeq10 \,T$, otherwise the neutrino burst from SN 1987A would have been shorter and with higher energies. Thus, we assume this saturation effect to model an unknown behavior of the spin fluctuation rate at very high densities encountered in the innermost regions of the SN core.

The adimensional part of Eq.~\eqref{eq:sigma} involving only nuclear physics is the function $s$ defined as
\begin{equation}
\begin{split}
        s(x_{a})&=s^{nn}(x_{a})+s^{pp}(x_{a})+s^{np}(x_{a})\,,\\
    s^{NN}&=\frac{1}{3}C_{aN}^{2}Y_{N}^{2}(s_{\bk}+s_{\bl}+s_{\bk\bl}-3s_{\bk\cdot\bl})\,,\\
s^{np}&=\frac{4}{3}Y_{n}Y_{p}\left[\left(C_{+}^{2}+C_{-}^{2}\right)s_{\bk}+\left(4C_{+}^{2}+2C_{-}^{2}\right)s_{\bl}\right]+\\
&-\frac{8}{3}Y_{n}Y_{p}\left[\left(C_{+}^{2}+C_{-}^{2}\right)s_{\bk\bl}-\left(3C_{+}^{2}+C_{-}^{2}\right)s_{\bk\cdot\bl}\right] \,\ ,    
    \end{split}
\label{eq:StructureFunc}
\end{equation} 
with $x_{a}=E_{a}/T$ and reflecting the structure of the matrix element in Eq.~\eqref{eq:MatElementBrem}.
For an arbitrary nucleon degeneracy, it is possible to write
\begin{widetext}
\begin{equation}
\begin{split}
s_{j}(x_{a})&=\int \frac{d\cos\delta}{2}\, \frac{d\phi}{2\pi} \,\frac{\sqrt{w}\,dw}{\sqrt{\pi}/2}\,du\,\frac{d\cos\theta}{2}\left[\frac{\rho Y_{1}}{2m_{N}}\left(\frac{2\pi}{m_{N}T}\right)^{1.5}\right]^{-1}\left[\frac{\rho Y_{2}}{2m_{N}}\left(\frac{2\pi}{m_{N}T}\right)^{1.5}\right]^{-1}\\
&\sqrt{u(u-x_{a})}e^{w-{\hat{\mu}}_{3}}e^{u-{\hat{\mu}}_{4}}H^{+}_{u}({\hat{\mu}}_{1})H^{-}_{u}({\hat{\mu}}_{2})H^{+}_{v}({\hat{\mu}}_{3})H^{-}_{v}({\hat{\mu}}_{4})\bigg|_{v=u-x_{a}\ge0}
\begin{cases}
    \tilde{\mathcal{F}}_{-}^{2}&{\rm if}\quad j=\bk\,,\\
    \tilde{\mathcal{F}}_{+}^{2}&{\rm if}\quad  j=\bl\,,\\
    \tilde{\mathcal{F}}_{+} \tilde{\mathcal{F}}_{-}&{\rm if}\quad  j=\bk\bl\,,\\
\frac{\xi}{3}\tilde{\mathcal{F}}_{+} \tilde{\mathcal{F}}_{-}&{\rm if} \quad j=\bk\cdot\bl\,,    
\end{cases}
\label{eq:snn}
\end{split}
\end{equation}
\end{widetext}
where the quantities labeled with numbers refer to nucleons in the initial state, 1 and 2, and final state, 3 and 4.
In Eq.~\eqref{eq:snn} we introduced the following functions derived from the nucleon Fermi-Dirac distributions
\begin{equation}
    \begin{split}
H^{\pm}_{u}({\hat{\mu}})&=(e^{\frac{w+u}{2}\pm\sqrt{uw}\cos\delta-{\hat{\mu}}}+1)^{-1} \,, \\
H^{\pm}_{v}({\hat{\mu}})&= (e^{\frac{w+v}{2}\pm\sqrt{vw}(\sin\delta\sin\theta\cos\phi+\cos\delta\cos\theta)-{\hat{\mu}}}+1)^{-1} \,.    
    \end{split}
\end{equation}
The axion spectrum in Eq.~\eqref{eq:dnNN} is evaluated by including effective nucleon masses $m_{N}^{*}$ through the replacement $m_{N}\to m_{N}^{*}$ is the formulas above. The same modification is applied also to the other emission processes discussed in the following.

\subsection{Pion-axion conversion}

In the recent literature it was shown that the pion-axion conversion, whose Feynman diagrams are shown in Fig.~\ref{fig:feynm2}, is another relevant axion production mechanism in SNe~\cite{Fischer:2021jfm,Carenza:2020cis}. The axion emission spectrum can be calculated as~\cite{Lella:2022uwi}
\begin{equation}
    \begin{split}
        &\left(\frac{d^2n_a}{dE_a\,dt}\right)_{N\pi}=\frac{g_{a}^2 T^{1.5}}{2^{1.5}\pi^5m_N^{0.5}}\left(\frac{g_A}{2f_\pi}\right)^2\left(E_a^2-m_a^2\right)^\frac{1}{2}\\
&\times\,\mathcal{C}_a^{p\pi^-}\frac{\Theta(E_{a}-\max\left(m_{a},m_{\pi}\right))}{\exp{\left(x_a-y_\pi-\hat{\mu}_\pi\right)}-1}\,(E_a^2-m_\pi^2)^\frac{1}{2}\frac{E_{a}^{2}}{E_{a}^{2}+\Gamma^{2}}\\
        &\times\int_0^\infty dy\, y^2\frac{1}{\exp{\left(y^2-\hat{\mu}_p\right)}+1}\frac{1}{\exp{\left(-y^2+\hat{\mu}_n\right)}+1}\,,
    \end{split}
    \label{eq:QaPionDef}
\end{equation}
where $y_{\pi}=m_{\pi}/T$, $\hat{\mu}$ is the degeneracy parameter for pions and nucleons, the Heaviside theta function fixes the minimal threshold energy and $\Gamma=0.5\Gamma_{\sigma}$ accounts for the multiple nucleon scattering effect. To be conservative, this is taken to be maximal given that the normalization condition in Eq.~\eqref{eq:normal} cannot be applied to this process~\cite{Raffelt:1993ix}.

The term related to the matrix element can be written as
\begin{equation}
    \mathcal{C}_a^{p\pi^-}=\frac{m_N^2}{g_A^2}\,\beta_a^2 \,\mathcal{G}_a(|\mathbf{p}_\pi|)\,,
    \label{eq:cpion}
\end{equation}
neglecting terms of higher order in $1/m_N$~\cite{Ho:2022oaw}
\begin{equation}
    \begin{split}
\mathcal{G}_a(|\mathbf{p}_\pi|)&=
\frac{2 g_{A}^{2}\big(2C_{+}^2 + C_{-}^2\big)}{3}
\bigg(\frac{|\bk_{\pi}|}{m_N}\bigg)^{2}
+C_{a\pi N}^{2}\bigg(\frac{E_{\pi}}{m_{N}}\bigg)^{2}+\\
&\quad+\frac{g_{A}^{2}C_{aN\Delta}^2 }{9}\mathcal{F}_{a\pi N}
\bigg(\frac{|\bk_{\pi}|}{m_N}\bigg)^{2} +\\
&\quad-\frac{4\sqrt{3} g_A^{2} C_{aN\Delta} }{9}
\mathcal{F}_{a N\Delta}
\bigg(\frac{|\bk_{\pi}|}{m_N}\bigg)^{2}\,,\\
\mathcal{D}&=\big[ 
\big(\Delta m - E_{\pi}\big)^{2} + \Gamma_\Delta^2 /4\big]\big[ 
\big(\Delta m + E_{\pi}\big)^{2} + \Gamma_\Delta^2 /4\big]\,,\\
\mathcal{F}_{a\pi N}&=\mathcal{D}^{-1}E_{\pi}^2 \Big(\Delta m^2 + 2 E_{\pi}^2 + \frac{\Gamma_\Delta^2}{4} \Big)
\,,\\
\mathcal{F}_{a N\Delta}&=\mathcal{D}^{-1}E_{\pi}
\Big[\big(\Delta m^2 - E_{\pi}^2 \big) \big(C_{+} \Delta m + C_{-} E_{\pi}\big) 
+ \\
&\quad\quad+
\frac{\Gamma_\Delta^2}{4} \big(C_{+} \Delta m - C_{-}E_{\pi} \big) \Big]\,,\\
    \end{split}
\end{equation}
where the width of the $\Delta$-resonance is $\Gamma_{\Delta}=117$~MeV, the nucleon-$\Delta$ mass difference is $\Delta m=m_{\Delta}-m_{N}^{*}$, with $m_{N}^{*}$ effective nucleon mass. 
Because of the energy conservation $E_{\pi}=E_{a}$ and Eq.~\eqref{eq:cpion} depends on the pion  momentum, that can be written in terms of the axion energy as $|\bp_{\pi}|^{2}=E_{a}^{2}-m_{\pi}^{2}$.
Similarly to the bremsstrahlung case, the squared matrix element for massive axions differs from the massless axion one for a factor $\beta_a^2$~\cite{Choi:2021ign,Ho:2022oaw,Lella:2022uwi}. 
In agreement with~\cite{Choi:2021ign,Ho:2022oaw}, the effect of the $\Delta$ resonance is to enahance the axion emissivity by a factor of $\sim2$ for $m_a\lesssim 500~\MeV$, depending on the chosen axion model.

\section{Modified luminosity criterion}
\label{sec:luminosity}
When Eqs.~\eqref{eq:dnNN}-\eqref{eq:QaPionDef} are integrated over a spherically symmetric SN model, we obtain the axion spectrum as~\cite{Chang:2018rso,Chang:2016ntp,Caputo:2021rux,Caputo:2022rca}  
\begin{equation}
    \frac{d^2N_a}{dE_a\,dt}=\int_0^\infty4\pi r^2 dr\left\langle e^{-\tau(E_a, r)}\right\rangle\,\frac{d^2 n_a}{dE_a\,dt}\,,    
\end{equation}
where $r$ is the radial coordinate from the center of the SN core and $\tau (E_a, r)$ is the optical depth at a given axion energy and location. 
The exponential term encodes the axion absorption effects in the SN and it is calculated by averaging over the cosine of the emission angle $\mu$ as~\cite{Caputo:2021rux}
\begin{equation}
    \left\langle e^{-\tau(E_a, r)}\right\rangle=\frac{1}{2}\int_{-1}^{+1}d\mu\,e^{-\int_0^\infty ds\, \lambda_a^{-1}\left(E_a\right)\big|_{\sqrt{r^2+s^2+2rs\mu}}}\,,
\end{equation}
to take into account the axion emission and propagation in any direction.
Here, the absorption axion mean-free path is calculated in the location indicated by the radial coordinate $\sqrt{r^2+s^2+2rs\mu}$ as~\cite{Giannotti:2005tn}
\begin{equation}    \lambda_{a}^{-1}(E_{a})=2\pi^2\left(E_a^2-m_a^2\right)^{-1}\,e^{\frac{E_{a}}{T}}\,\frac{d^{2}n_a}{dE_a\,dt}(\chi\, E_{a}) \,\ ,
\end{equation}
where $\chi=\pm1$ for pion-axion conversion  and bremsstrahlung, respectively. Moreover, for the pionic absorption the phase space changes because of a pion in the final state, requiring the substitution of the distribution function $f_\pi\rightarrow1+f_\pi$. 

Axions are said to be in the free-streaming regime when this factor does not differ too much from one in any region of the SN. In the opposite case, axions are in the trapping regime and their emission is typically suppressed in the innermost regions of the star. Their emission occurs from a relatively thin spherical surface, called axionsphere. In this case, their spectrum is mostly determined by the properties of the SN plasma at the location of the axionsphere.
\begin{figure} [t!]
\centering
    \includegraphics[width=1\columnwidth]{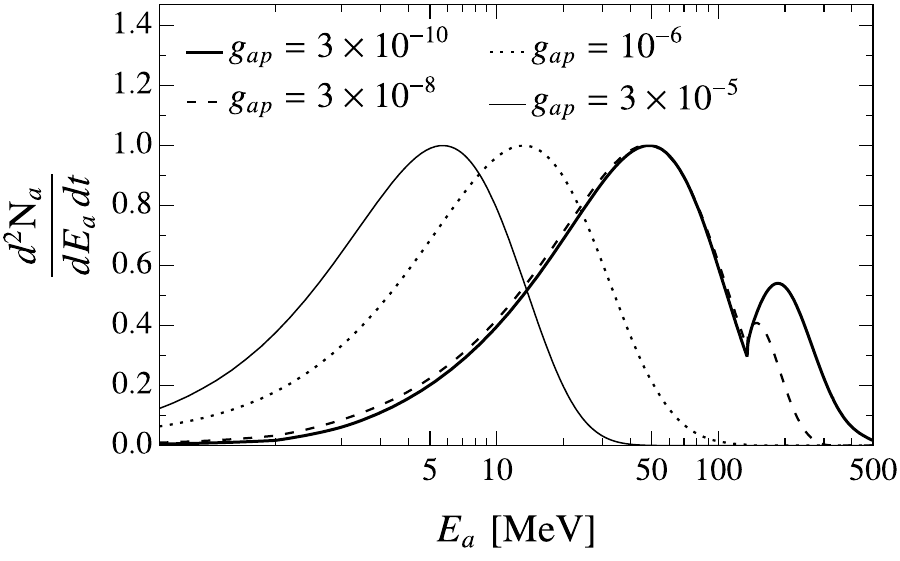}
    \caption{Axion spectra in the massless case at $t_{\rm pb}=1\,\s$ for various axion-proton couplings. The spectra are arbitrarily normalized for visualization purposes. Figure taken from~\cite{Lella:2023bfb}.}
	\label{fig: Spectra}
\end{figure}
These integrations are performed over a 1D spherically symmetric and general relativistic hydrodynamics model of an $18~M_{\odot}$ progenitor, based on the {\tt AGILE BOLTZTRAN} code~\cite{Mezzacappa:1993gn,Liebendoerfer:2002xn}. Furthermore, the pion chemical potentials and a pion abundances are included in the simulation following~\cite{Fore:2019wib,Fischer:2021jfm}.

Figure~\ref{fig: Spectra} shows the behaviour of axion spectra for light axions and different values of the axion-proton coupling $g_{ap}$. In the free streaming regime, $g_{ap}=3\times10^{-10}$ (solid thick line), the spectrum features two peaks, one at ${E_a\simeq 50}$~MeV associated to the bremsstrahlung emission, and the other one at $E_a \simeq 200$~MeV due to pionic processes~\cite{Lella:2022uwi,Lella:2023bfb}. As the coupling is increased we enter the trapping regime.
Axion absorption happens via bremsstrahlung $aNN\to NN$ for relatively low energies, $E_{a}\lesssim m_{\pi}$. At higher energies, pionic processes $aN\to \pi N$  dominate the absorption~\cite{Lella:2023bfb}.
The stronger absorption in the part of the spectrum at higher energies is clear for $g_{ap}=3\times10^{-8}$ (dashed line), where the high-energy peak is strongly suppressed. Increasing the coupling even more, it disappears completely as evident from the case $g_{ap}=10^{-6}$ (dotted line) and $g_{ap}=3\times10^{-5}$ (solid thin curve).
The bremsstrahlung spectrum, with energies around $50$~MeV in the free-streaming regime, becomes colder in the trapping regime, down to a few MeV of energy.  
Indeed, the axion spectrum reflects the temperature of the regions where escaping axions are produced. In the free-streaming regime, it corresponds to the entire SN core and the spectrum depends on average properties of the core. 
As the axion-nucleon coupling increases, axions become trapped in the innermost regions of the SN core and only the ones produced in the outer SN layers, where the temperature decreases, are able of escaping. 
Thus, the average energy of the spectrum decreases in the trapping regime.

\begin{figure} [t!]
\centering
    \includegraphics[width=1\columnwidth]{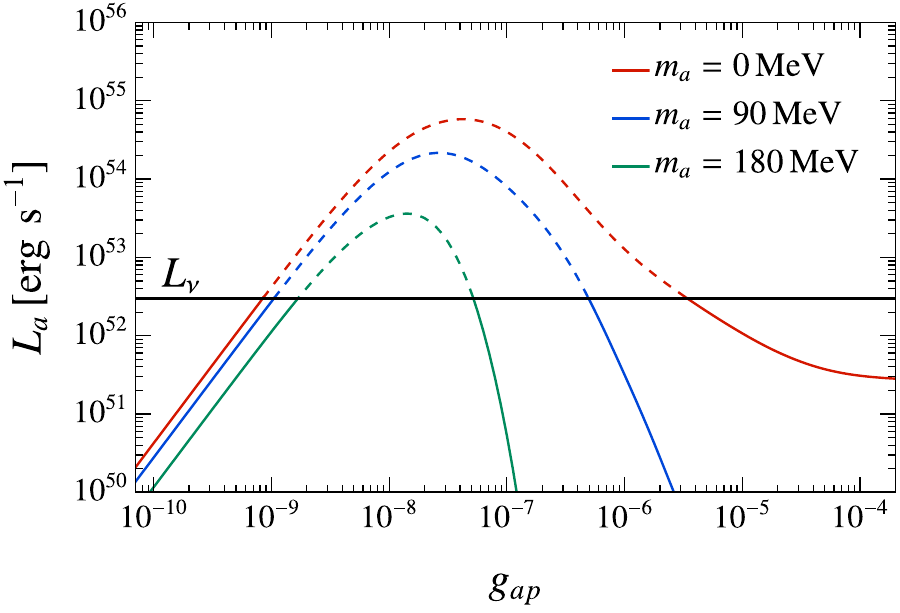}
    \caption{Axion luminosity at $t_{\rm pb}=1\,\s$ as function of the axion-proton coupling $g_{ap}$ for three different axion masses and the black line indicates the value of the neutrino luminosity. When the axion luminosity exceeds this limit, we used dashed lines to indicate the lines.
    Figure taken from~\cite{Lella:2023bfb}.}
	\label{fig:lavsgap}
\end{figure}

A quantity that characterizes the amount of energy stolen by axions from the SN is the luminosity, defined as~\cite{Caputo:2021rux,Chang:2016ntp,Lucente:2020whw}
\begin{equation}
    L_a=\int_0^\infty4\pi r^2 dr\int_{m_a/\alpha}^\infty dE_a\, E_a \, \alpha(r)^2 \left\langle e^{-\tau(E_a, r)}\right\rangle\,\frac{d^2 n_a}{dE_a\,dt}\,,    
\end{equation}
where $\alpha$ is the lapse factor, accounting for gravitational redshift, entering also in the lower limit of integration $m_a/\alpha$, which cuts away the fraction of heavy axions gravitationally trapped in the interior of the core~\cite{Caputo:2022mah,Lella:2022uwi,Lucente:2020whw}. In Fig.~\ref{fig:lavsgap} we show the axion luminosity as function of the axion-proton coupling. For sufficiently small coupling, $g_{ap}\lesssim 10^{-8}$, the luminosity increases quadratically with $g_{ap}$. This is a property of the free-streaming regime, where the axion production increases because of the stronger interactions with nuclear matter. For higher couplings, absorption processes become relevant and lower the mean energy of emitted axions, as we discussed. In this case, in the trapping regime, the luminosity drops. In Fig.~\ref{fig:lavsgap} we show this behavior for different axion masses and heavier axions suffer more the suppression in the trapping regime because their relevant interactions are the very efficient pionic processes. Thus, their absorption is more efficient and the production is Boltzmann suppressed.

\begin{figure*}[t!]
\vspace{0.cm}
\includegraphics[scale=0.95]{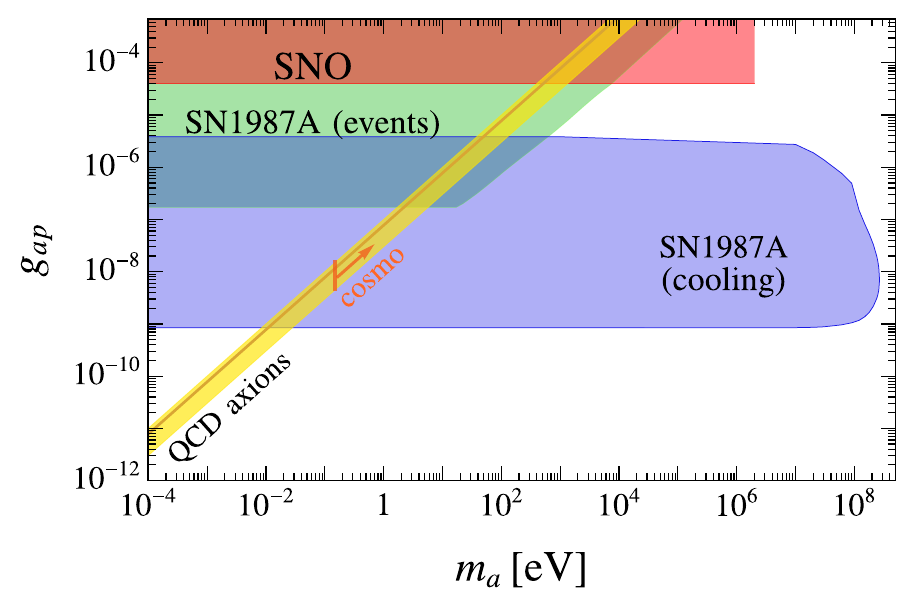}
\caption{Summary plot of the bounds for axions coupled to protons.  The QCD axion band is shown in yellow. 
The region labeled SNO is excluded by the search of solar axions dissociating deuteron~\cite{Bhusal:2020bvx}.  
The green region is constrained by the absence of an axion excess in KII in coincidence with the neutrino burst.
The blue region is excluded by the cooling argument applied to SN 1987A. 
The orange line with the arrow within the QCD axion band shows the reach of cosmological experiments in the next future, $m_a\gtrsim0.15~\eV$~\cite{Archidiacono:2015mda,DEramo:2022nvb}. Figure taken from~\cite{Lella:2023bfb}.}
\label{fig:CoolingBound}
\end{figure*}

\section{Cooling bound}
\label{sec:cooling}
Axions produced in a SN escape from the star stealing energy from the core. In the free-streaming regime, their interactions with ordinary matter in the stellar medium are weak and, once produced, they do not interact anymore in the SN, efficiently draining energy.
Deeply in the trapping regime, axions interact strongly with the stellar medium and their energy is partially redeposited inside the star in their interactions with the SN matter. This feature is connected with the decrease in the average energy as the axion-nucleon coupling is increased.
When the energy released by the SN in form of axions is comparable with the energy drained by neutrino during the SN cooling phase, axions are expected to significantly affect the SN evolution.
Therefore, our knowledge of SN physics, mostly corroborated by the observation of SN 1987A, can be used to constrain axions stealing a large amount energy from the star.
If the axion luminosity $L_{a}$ is larger than the neutrino one $L_{\nu}$, considering neutrinos of all flavors, the SN neutrino burst would be significantly shorter than what observed from SN 1987A.
This criterion is used to constrain axion properties, as it is widely applicable to several models of exotic particles with weak interaction. The precise formulation of this constraint is~\cite{Raffelt:1990yz}
\begin{equation}
    L_{a}\lesssim3\times10^{52}\,{\rm erg}{\rm s}^{-1}\,,
\end{equation}
at the beginning of the cooling phase, at about $t_{\rm pb}=1$~s, where $t_{\rm pb}$ is the post-bounce time.

Applying this criterion to the axion case, we can exclude the blue region in  Fig.~\ref{fig:CoolingBound}~\cite{Lella:2023bfb}. Precisely, it is possible to constrain axions with
\mbox{$8\times10^{-10}\lesssim g_{ap}\lesssim 4\times10^{-6}$} for \mbox{$m_a\lesssim 10\MeV$} and \mbox{$10^{-9}\lesssim g_{ap}\lesssim 1.5\times10^{-7}$} for \mbox{$m_a\sim\mathcal{O}(100)\MeV$}. For KSVZ axions, the constraint reads $m_{a}\lesssim11$~meV.
The lower end of the bound corresponds to the free-streaming regime, for lower couplings not enough axions are produced in the SN to compete with neutrinos in subtracting energy from the core. The upper end of the bound falls in the trapping regime. Here, axions with couplings $g_{ap} \gtrsim 4\times10^{-6}$, would re-deposit energy in the star on their way out because of their strong interactions with the SN matter. At this point it is worth to stress that the discussed constraints are obtained for KSVZ axions assuming $C_{an}=0$. If we consider $C_{an}=-0.02$~\cite{GrillidiCortona:2015jxo}, the axion luminosity increases of only a few percents. For other axion models where $|C_{an}|\simeq |C_{ap}|$, the showed luminosity can increase of a factor between several percents and two, depending on the model. 

In Fig.~\ref{fig:CoolingBound} we show other constraints in this parameter space. The green bound excludes axions that would have been produced in SN 1987A and interact in Kamiokande-II via nuclear absorption $a+\,^{16}{\rm O}\to\,^{16}{\rm O}^{*}$, giving a signature indistinguishable from the neutrino one in the oxygen radiative de-excitation. Thus, the non-observation of a similar signal sets constraints on strongly interacting axions~\cite{Lella:2023bfb,Carenza:2023wsm}. The red constraint excludes the possibility that axions produced in the Sun in the process of formation of a helium isotope by the proton-deuteron scattering, $p+d\to\,^{3}{\rm He}+a(5.5~\MeV)$, are detected in the Sudbury Neutrino Observatory by dissociation of deuteron in the heavy water of the detector~\cite{Bhusal:2020bvx}. These constraints are compared with the yellow band of most common QCD axion models. The orange arrow denotes the reach of cosmological experiments, that in the next future will be able to probe axions with $m_{a}\gtrsim0.15~\eV$~\cite{Archidiacono:2015mda,DEramo:2022nvb}. From Fig.~\ref{fig:CoolingBound} it is clear that SN constraints, based on different reasoning, are complementary to laboratory and cosmological searches.

\section{Conclusions}
\label{sec:conclusions}

Supernovae offer captivating opportunities to investigate the properties and interactions of various particles, presenting a unique window into fundamental physics. This study has specifically explored the recent progress made in understanding axion emission from hot and dense nuclear matter in SNe. By focusing on the main production mechanisms for axions coupled with nucleons, namely bremsstrahlung and pion-axion conversion, significant insights have been gained into the behavior of axions in these extreme astrophysical environments.

The characterization of SN axion emission is of great importance for our understanding of axions and their properties. By delving into the mechanisms of axion production, this work is a guideline to evaluate the state-of-the-art SN axion fluxes and study their phenomenology.

\begin{acknowledgments}
A special thank to Alessandro Lella for helpful comments.
This article is based upon work from COST Action COSMIC WISPers CA21106, supported by COST (European Cooperation in Science and Technology).
The work of P.C. is supported by the European Research Council under Grant No.~742104 and by the Swedish Research Council (VR) under grants  2018-03641 and 2019-02337. 
\end{acknowledgments}

\section*{Data Availability Statement} No data associated with the manuscript.

\bibliographystyle{bibi}
\bibliography{references.bib}

\end{document}